\documentclass[preprint,5p,times,twocolumn]{elsarticle}
\biboptions{sort&compress}
\usepackage{graphicx} 
\usepackage{float}
\usepackage{amssymb}
\usepackage{lipsum}  
\usepackage{placeins}
\usepackage{subcaption}
\usepackage{overpic}
\usepackage{amsmath}
\usepackage{url}
\usepackage{gensymb}
\usepackage{ulem}
\usepackage[pdftex, pdftitle={Article}, pdfauthor={Author},colorlinks = true,allcolors = blue]{hyperref}
\usepackage{soul}
\usepackage{calrsfs}
\DeclareMathAlphabet{\pazocal}{OMS}{zplm}{m}{n}

\usepackage{lineno}
\begin{document}

\title{Measurement of SiPM Dark Currents and Annealing Recovery \\for Fluences Expected in ePIC Calorimeters at the Electron-Ion Collider}

\author[inst1]{Jiajun Huang}
\author[inst1]{Sean Preins}
\author[inst1]{Ryan Tsiao}
\author[inst1]{Miguel Rodriguez}
\author[inst1]{Barak Schmookler\fnref{fn2}}
\author[inst1]{Miguel Arratia\fnref{fn1}}

\address[inst1]{Department of Physics and Astronomy, University of California, Riverside, CA 92521, USA}

\fntext[fn1]{corresponding author, miguel.arratia@ucr.edu}
\fntext[fn2]{Now at University of Houston.}

\begin{abstract}
Silicon photomultipliers (SiPMs) will be used to read out all calorimeters in the ePIC experiment at the Electron-Ion Collider (EIC). A thorough characterization of the radiation damage expected for SiPMs under anticipated EIC fluences is essential for accurate simulations, detector design, and effective operational strategies. In this study, we evaluate radiation damage for the specific SiPM models chosen for ePIC across the complete fluence range anticipated at the EIC, $10^8$ to $10^{12}$ 1-MeV $n_{\mathrm{eq}}$/cm$^2$ per year, depending on the calorimeter location. The SiPMs were irradiated using a 64 MeV proton beam provided by the University of California, Davis 76" Cyclotron. We measured the SiPM dark-current as a function of fluence and bias voltage and investigated the effectiveness of high-temperature annealing to recover radiation damage. These results provide a comprehensive reference for the design, simulation, and operational planning of all ePIC calorimeter systems.
\end{abstract}

\maketitle

\section{Introduction}
Silicon photomultipliers (SiPMs) are increasingly used in particle and nuclear physics experiments~\cite{Simon:2018xzl}. The ePIC detector at the future Electron-Ion Collider~\cite{Accardi:2012qut,ABDULKHALEK2022122447}, incorporating concepts from Refs.\cite{Adkins:2022jfp,ATHENA:2022hxb}, will be among the first collider experiments using SiPMs from day one. Approximately $\mathcal{O}(1)$ million SiPMs will be used to read out all calorimeter systems~\cite{Bock:2022lwp}, as well as additional subsystems such as Cherenkov imaging detectors.

While SiPMs offer numerous advantages over traditional photosensor technologies~\cite{Simon:2018xzl}, their susceptibility to radiation damage~\cite{Garutti:2018hfu} must be thoroughly understood. Such understanding is critical to inform detector design, perform realistic simulations, and develop robust operational strategies for experiments at colliders, particularly those using proton or nuclear beams like the EIC.

The nominal luminosity of the EIC depends on the operating center-of-mass energy, peaking at $10^{34}$ cm$^{-2}$s$^{-1}$\cite{ABDULKHALEK2022122447}. Consequently, calorimeters will experience annual fluences between $10^8$ and $10^{12}$ 1-MeV neutron equivalent per cm$^{2}$ ($n_{\mathrm{eq}}$/cm$^2$), depending on their location~\cite{ePICFluence}. Although this fluence generally produces moderate damage to SiPMs, characterized mainly by increased dark current rather than other performance degradations~\cite{Garutti:2018hfu}, the damage is amplified by the anticipated room-temperature operating conditions of most ePIC calorimeters. Dark current increases noise and may become problematic for system operation and temperature control, as well as affect the calorimeter cell thresholds, ultimately degrading physics performance.

Several previous studies in the literature have examined SiPM radiation damage (see, e.g., Ref.\cite{Garutti:2018hfu}), yet a detailed, model-specific characterization across the complete EIC-relevant fluence range is still lacking. Additionally, it is important to understand the potential recovery of SiPM performance through high-temperature annealing\cite{Garutti:2018hfu}, a factor crucial for detectors subjected to the highest radiation levels.

At the EIC, the highest fluences ($10^{11}$ to $10^{12}$ 1-MeV $n_{\mathrm{eq}}$/cm$^2$ per year) are expected for subdetectors nearest the beam pipe along the hadron-beam direction, specifically the innermost section ($r<50$ cm) of the forward HCAL~\cite{Bock:2022lwp}, the calorimeter Insert~\cite{Arratia:2022quz}, and the Zero-degree calorimeter~\cite{Milton:2024bqv}. All these subsystems will operate at room temperature. By comparison, the SiPM-on-tile section of the CMS High-Granularity Calorimeter will be exposed to a maximum fluence of approximately $8\times 10^{13}$ 1-MeV $n_{\mathrm{eq}}$/cm$^{2}$ over its 10-year lifetime, but it operates at $-30^\circ$C to suppress dark current by a factor of 30~\cite{CMS:2017jpq}.

Consequently, one year of CMS HGCAL operation at the HL-LHC will yield dark current levels roughly equivalent to those experienced by the aforementioned EIC calorimeters. In either case, operating these detectors will likely push the limits regarding noise levels, underscoring the need for accurate measurements of the expected dark current and the recovery achieved through annealing, which is a mitigation option built into the design of the calorimeter insert~\cite{Arratia:2022quz} and the ZDC~\cite{Milton:2024bqv}.

This work provides a comprehensive, unified reference on SiPM dark current induced by radiation damage relevant to the ePIC calorimeter systems. Section~\ref{sec:setup} describes the experimental setup, Section~\ref{sec:results} presents the results, and Section~\ref{sec:conclusion} summarizes the conclusions.
\section{Experimental Setup}
\label{sec:setup}
\subsection{Irradiation facility and SiPM models tested}
\label{sec:facility}
All SiPMs studied in this work were irradiated at the University of California, Davis, using the 76-inch Isochronous Cyclotron, which delivers a proton beam with a kinetic energy of 64 MeV. The cyclotron beam fluence is relatively uniform within a radial distribution of 3 cm~\cite{DavisCyclotron}. The fluence is measured to better than 10$\%$ with an ionization chamber. For reference, a 64 MeV proton beam induces approximately 50\% more damage compared to a 1 MeV neutron beam~\cite{AKKERMAN2001301}.
\begin{equation}
    \frac{K_{64 MeV p^+}}{K_{1 MeV n}} = \frac{NIEL_{64 MeV p^+}}{NIEL_{1 MeV n}} = \frac{\Phi_{1 MeV n}}{\Phi_{64 MeV p^+}} \approx 1.5 \thinspace,
\end{equation}
where $K$ is the damage factor, $NIEL$ is the non-ionizing energy loss, and $\Phi$ is the equivalent fluence.

\begin{figure}[h!]
    \centering
     \includegraphics[width=0.8\linewidth]{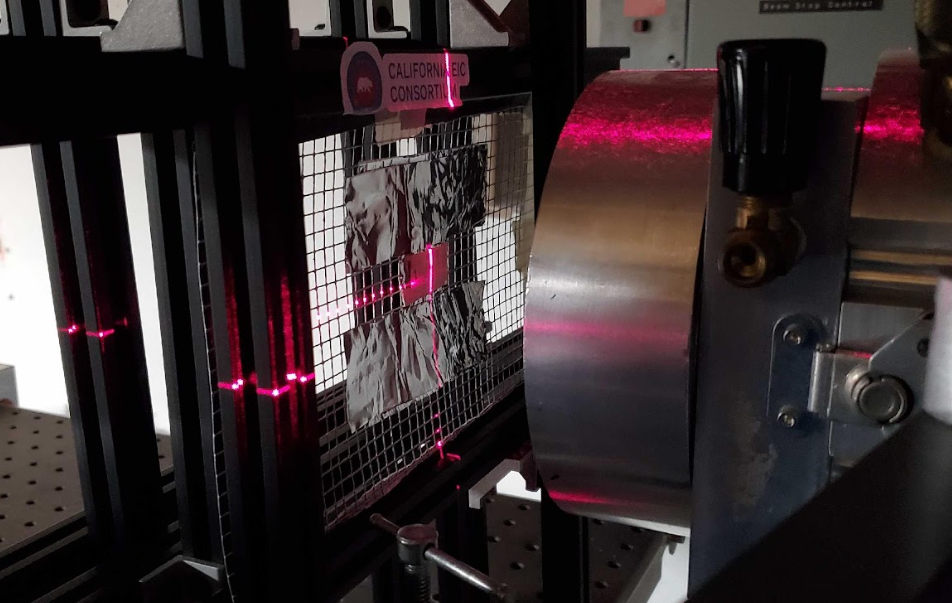} 
         \includegraphics[width=0.34\textwidth]{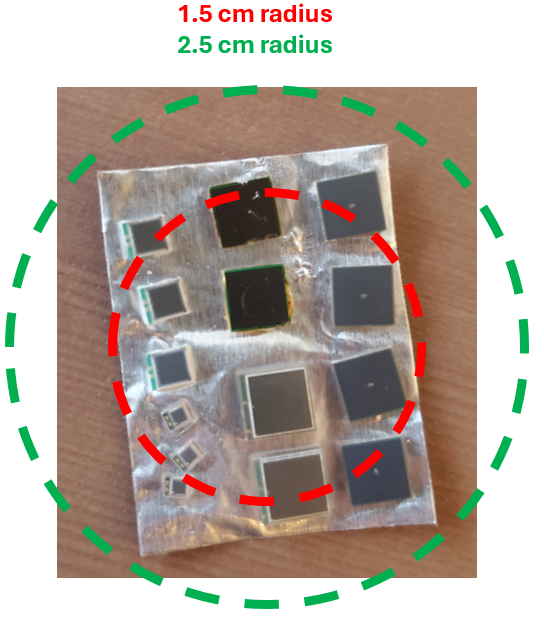} 
    \caption{Figure (a) (top) illustrates the laser alignment of the SiPM tape placed on the mesh, centered directly in front of the cyclotron beam exit. Figure (b) (bottom) shows various types of SiPMs positioned within a 3 cm radius for irradiation.}%
    \label{fig:beam_setup}%
\end{figure}

The SiPM models irradiated and characterized in this study are manufactured by Hamamatsu: S14160-6050HS, S14160-6015PS, S14160-3015PS, S14160-1315PS, and S13360-6050VE. A metal stand holding a metal mesh covering the beam exit window was used, with the SiPMs attached to the mesh by aluminum tape. This allowed the SiPMs to be positioned approximately 15 cm from the center of the exit window, as shown in \hyperref[fig:3015pre]{Fig.~\ref{fig:beam_setup}}. The tape containing a group of SiPMs was replaced after irradiation at each of the fluence levels: $10^8$, $10^9$, $10^{10}$, $10^{11}$, $10^{12}$, and $10^{13}$ protons/cm$^2$.

The number of units irradiated for each model at a given fluence point as well as the corresponding ePIC calorimeter is detailed in Table~\ref{tab:irradiation_table}.

\begin{table*}[h!]
\centering
\begin{tabular}{|l|c|c|c|c|c|c|c|}
\hline
\textbf{Models of SiPMs} & \textbf{$10^8$ N$_{p^+}$} & \textbf{$10^9$ N$_{p^+}$} & \textbf{$10^{10}$ N$_{p^+}$} & \textbf{$10^{11}$ N$_{p^+}$} & \textbf{$10^{12}$ N$_{p^+}$} & \textbf{$10^{13}$ N$_{p^+}$} & \textbf{ePIC Detector Usage} \\ \hline
S14160 1315PS           & 1                      & 3                      & 3                        & 3                        & 3                        & 2              &  nHCAL, pHCal         \\ \hline
S14160 3015PS           & 1                      & 2                      & 2                        & 3                        & 3                        & 1              & nEMCAL , bHCAL, pHCal(Insert), ZDC         \\ \hline
S14160 6015PS           & 1                      & 1                      & 1                        & 2                        & 2                        & 1              & nEMCAL, bEMCAL, pEMCAL          \\ \hline
S14160 6050HS           & 2                      & 4                      & 4                        & 4                        & 4                        & 2              & bEMCAL , pEMCAL, pHCAL (Insert), ZDC       \\ \hline
S13360 6050VE           & 2                      & 2                      & 2                        & 2                        & 2                        & 0              & bEMCAL          \\ \hline
\end{tabular}
\caption{Number of SiPMs irradiated at each fluence level. All SiPM models are manufactured by Hamamatsu. The calorimeter nomenclature is as follows: nHCAL, bHCAL, and pHCAL denote the electron-going endcap, barrel, and hadron-going endcap hadronic calorimeters, respectively. Similarly, nEMCAL, bEMCAL, and pEMCAL refer to the electron-going endcap, barrel, and hadron-going endcap electromagnetic calorimeters. ZDC stands for Zero Degree Calorimeter. The term ``Insert" refers to the region of the pHCAL covering the pseudorapidity range $3.0<\eta<4.0$ with respect to the proton beam.}
\label{tab:irradiation_table}
\end{table*}

\subsection{Characterization Setup}
\begin{figure}[h!]
    \centering
    \includegraphics[width=0.32\textwidth]{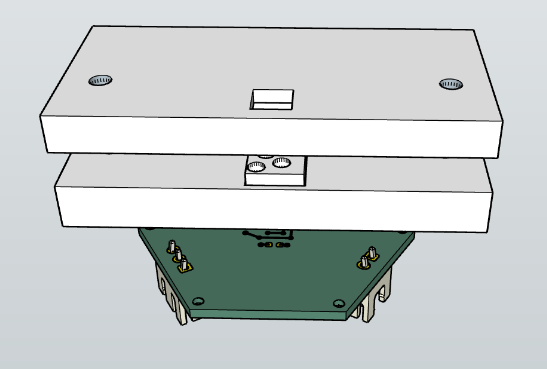} 
     \includegraphics[width=0.65\linewidth]{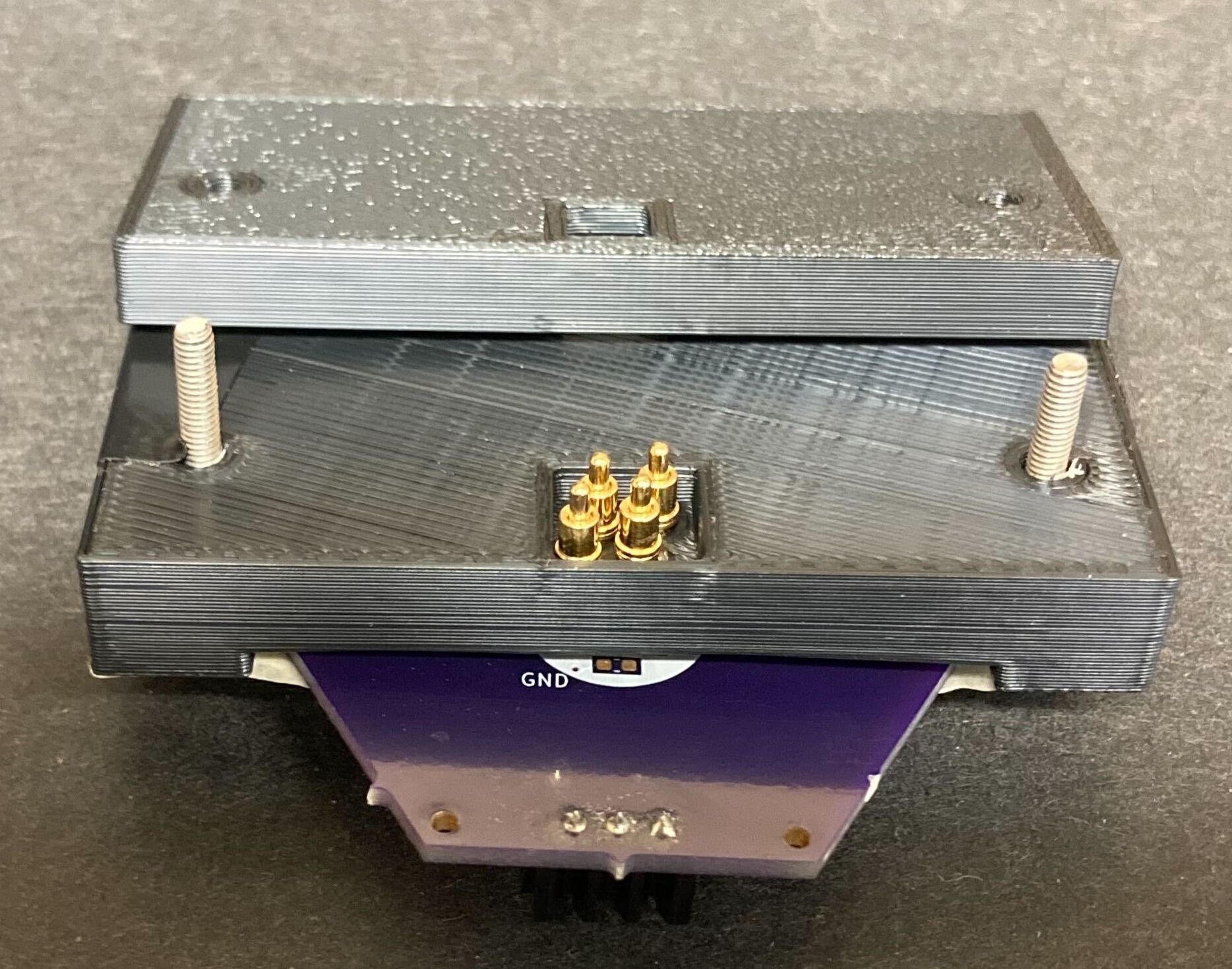} 
    \caption{Figure (a) (top) shows a SketchUp 3D model of the holders. Figure (b) (bottom) shows the 3D-printed holders with a hexagon board and pogo pins attached, ready for electrical connection.}%
    \label{fig:3d_model}%
\end{figure}

To prevent unintended annealing caused by soldering SiPMs directly onto measurement devices, a contact-based electrical connection setup using pogo pins was implemented, as illustrated in \hyperref[fig:3015pre]{Fig.~\ref{fig:3d_model}}. This setup features a hexagonal PCB designed to connect SiPM anodes and cathodes via pogo pins. It is secured with 3D-printed holders, and a custom cable with BNC and 3-pin connectors is used to interface with the board. The arrangement enables stable electrical connections and facilitates rapid switching between SiPMs without soldering. This setup was placed inside a dark box. 

The bias and readout instrument used was a Keithley 2450 SourceMeter, controlled automatically through LabVIEW to set voltages and measure dark current. Python code was employed for data processing and plotting. Although the Keithley 2450 can measure dark currents up to 100 mA at a bias voltage of +20 V~\cite{Keithley}, we limited the maximum measured dark current to 10 mA via software. 

The breakdown voltage, $V_{break}$, of each SiPM was estimated from its dark current, $I_{dark}$, measurements using Eq.~\ref{eq:breakdown} and was accounted for when calculating the overvoltage, defined as $V_{over} = V_{bias} - V_{break}$.
\begin{equation} \label{eq:breakdown}
    V_{break} = V_{applied}[max(\frac{\Delta log(I_{dark})}{\Delta log(V_{applied})})]
\end{equation}

\subsection{SiPM Storage and Annealing Setup}
All irradiated SiPMs were stored in a box at room temperature, and each SiPM was packed with 3D printed holders for individual spacing and identification purposes. All SiPMs were then transported from Davis, California back to Riverside, California after the radiation cool down period.

We estimate the impact of room-temperature annealing by comparing measurements taken at UC Davis, UC Riverside, and during an in-situ SiPM irradiation campaign at RHIC~\cite{Zhang:2025ijg}. Our model is $y = ae^{-bt} + c$, with $a = 0.209$, $b = 0.019$, $c = 0.768$, where $y$ represents the normalized dark current and $t$ the time in months. Using this model we obtain a correction factor for dark currents of $1.24 \pm 0.05$ after three months (the uncertainty reflects overvoltage dependence). This normalization uncertainty is the dominant systematic uncertainty on the dark-current measurements. 

An environmental control chamber (LIB TH-50)\cite{LIB} was used to perform constant high-temperature annealing of the irradiated SiPMs. A temperature of 120 $\degree$C was chosen as a trade-off between achieving a fast annealing process and avoiding damage that may occur at higher temperatures. During high-temperature annealing, the chamber was heated at a rate of $+3^\circ\text{C/min}$ until it reached the set temperature, which was then maintained for a specified duration. Afterwards, the chamber was cooled at a rate of $-1^\circ\text{C/min}$ until it reached a preset room temperature of $25^\circ\text{C}$. Only the duration of the constant-temperature phase, which ranged from hours to hundreds of hours, is included in the total annealing time reported in Section~\ref{sec:sec:postannealing}. After removal from the chamber, each SiPM was allowed to rest at room temperature until it reached thermal equilibrium with the ambient environment before data collection. Since the impact of room-temperature annealing is negligible compared to that of high-temperature annealing, no correction is applied.

\section{Results}
\label{sec:results}
\subsection{Pre-Annealing}
\begin{figure}[h!]
    \centering
   \includegraphics[width=0.825\linewidth]{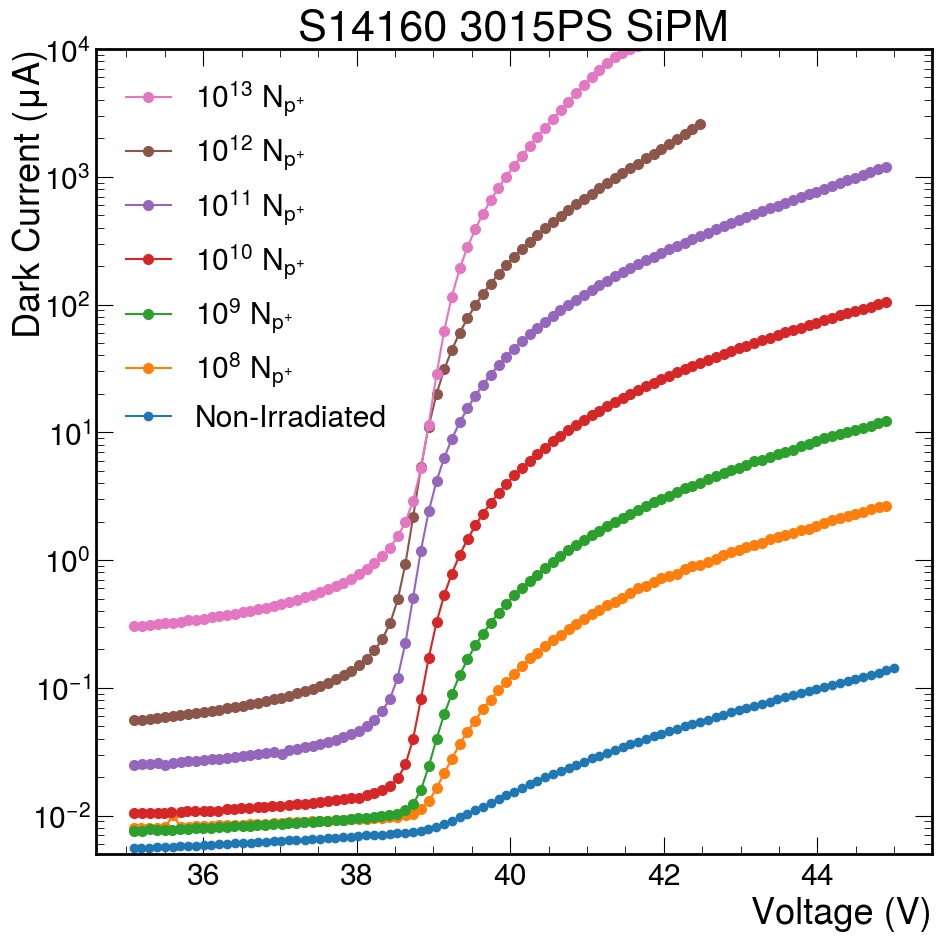}\\
   \includegraphics[width=0.825\linewidth]{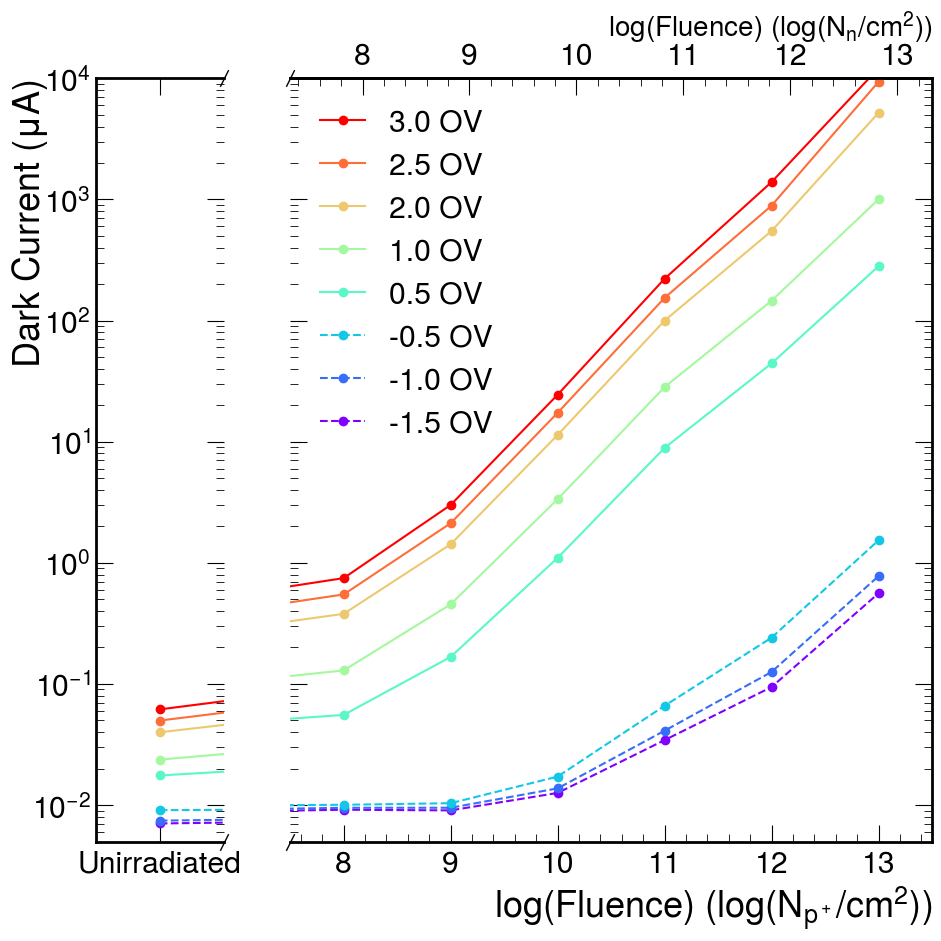} 
    \caption{Figure (a) (top) shows the relationship between dark current and applied voltage for S14160 3015 SiPMs tested at various fluences, compared with that of a non-irradiated model. Each curve represents a different SiPM unit, each of which may exhibit a slightly different overvoltage due to typical variations in this parameter.  Figure (b) (bottom) presents a logarithmic plot of fluence versus dark current at different overvoltages. In both plots, the data are scaled based on the acquisition date of each data point to account for room-temperature annealing effects. Measurements were taken at room temperature.}%
    \label{fig:3015pre}%
\end{figure}

Figure~\ref{fig:3015pre} shows the IV measurements for S14160 3015PS SiPMs recorded at various fluences and overvoltages, compared to an unirradiated reference. IV characterization plots for other SiPM types or families are provided in the Appendix, and the corresponding data are available on Zenodo~\cite{Data}. The dark current for positive overvoltages increases linearly with fluence, in agreement with previous studies (e.g., Refs.~\cite{Garutti:2018hfu,Altamura:2021lio}). For negative overvoltage values, the dark current becomes significantly higher than the uniarradiated reference only at fluences exceeding 10$^9$ 64‐MeV N$_{p^+}$.

\begin{figure}[h!]
    \centering
     \includegraphics[width=0.825\linewidth]{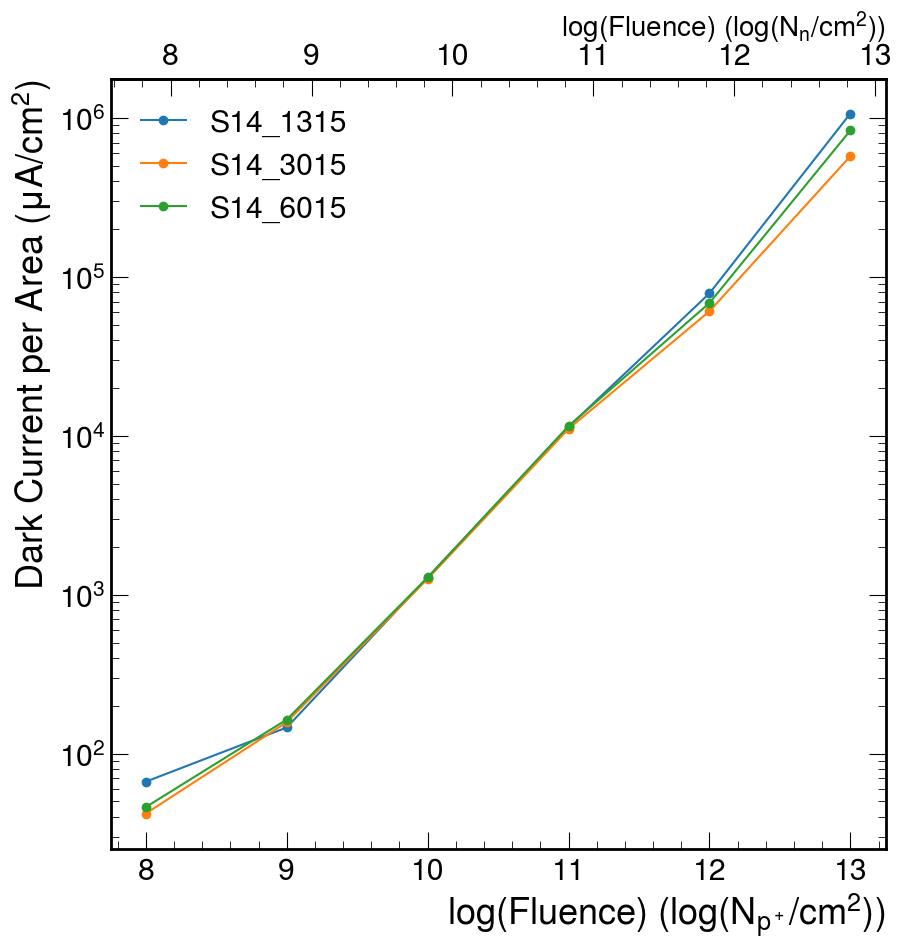}
    \caption{Dark current per unit area vs fluence for different types of SiPMs operating at +2V overvoltage. The data in the plot are scaled based on the acquisition date of each data point to account for room-temperature annealing effects.}
    \label{fig:compare}
\end{figure}

Figure~\ref{fig:compare} compares the dark current at an overvoltage of +2 V for different SiPM models with a pixel size of 15 $\mu$m but varying areas. When normalized by area, the measurements across all models are largely consistent, as expected. This consistency helps validate the measurements and constrain systematic effects arising from differences in absolute currents. 

\subsection{Post-Annealing}
\label{sec:sec:postannealing}
\begin{figure}[h!]
    \centering
    \includegraphics[width=0.825\linewidth]{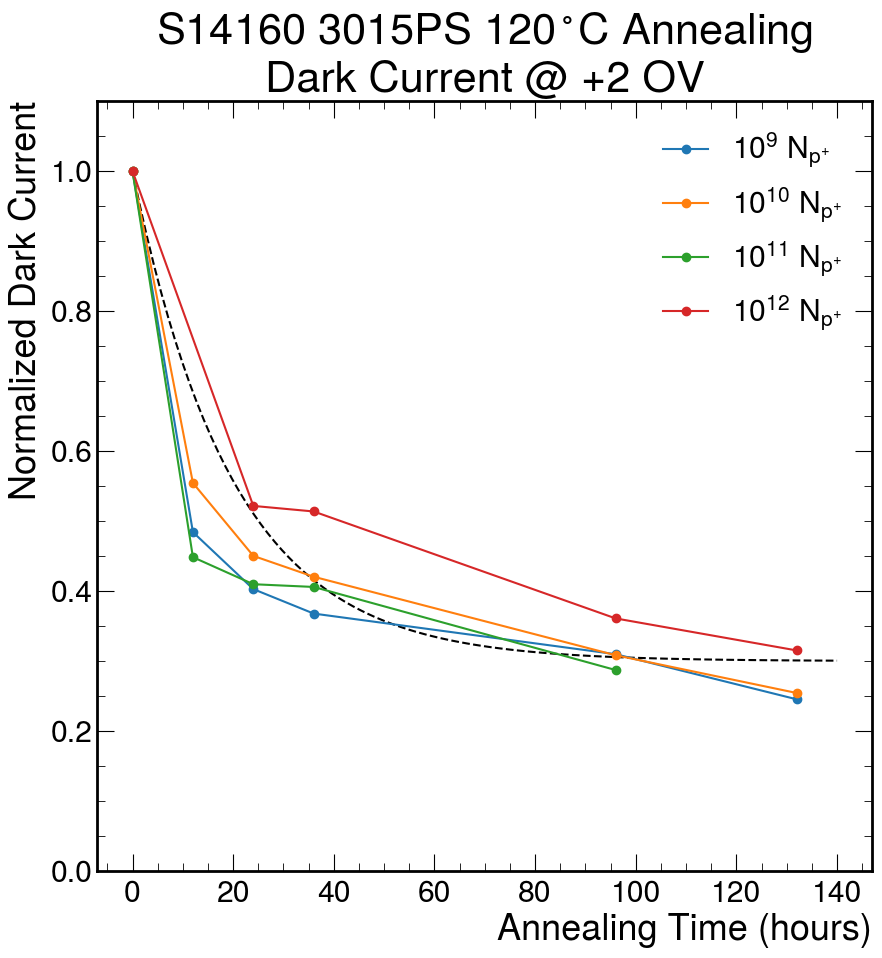}
    \caption{Dark current trends for various irradiation fluences measured at different annealing times at a constant temperature of 120 $\degree$C and at +2V overvoltages.}
    \label{fig:fluence_post}
\end{figure}

Figure~\ref{fig:fluence_post} compares the relative decrease in dark current as a function of annealing time at $120^\circ\text{C}$ for various fluences using the S14160-3015PS model. In each case, the dark current is normalized to the value measured before annealing. The data exhibit a similar trend across all fluences, following the expected exponential decay. On average, the initial dark current value is reduced by half after about 19 hours of annealing.

\FloatBarrier

\section{Summary and Outlook}
\label{sec:conclusion}
We have presented a comprehensive dataset of dark current in radiation-damaged SiPMs irradiated at fluence levels expected at the EIC. Our dataset comprises measurements spanning all relevant bias voltage ranges, fluence levels, and SiPM models used in every calorimeter subsystem of the ePIC detector. We also measured the recovery of radiation damage after high-temperature annealing—a process likely required for subsystems operating under the harshest conditions, such as the calorimeter Insert and the Zero-Degree Calorimeter.

Our dataset will underpin future studies on expected noise levels and realistic thresholds, which vary by system because different subsystems employ distinct digitization schemes ranging from ASICs to custom solutions with discrete components.

Our dataset informs realistic simulations and defines operational strategies, including determining the appropriate overvoltage for each subsystem, establishing mitigation strategies (such as planning high-temperature annealing), and setting safety factors. It will also enable realistic simulations of physics performance under the expected noise levels.

\section*{Data Availability}
The raw IV curve spectrums are available for access in Zenodo \cite{Data}, without room temperature annealing factors applied for the respected date that the data were taken. The data recording timestamps are noted in the file name.

\section*{Contributions}
J. Huang analyzed the data, developed LabVIEW scripts, and performed IV measurements at UC Riverside and UC Davis. He also edited the manuscript and produced the figures. S. Preins contributed to beam test preparation, running the experiment, and conducted measurements at UC Davis. R. Tsiao carried out measurements and data analysis at UC Davis. M. Rodriguez conducted annealing studies and performed IV measurements at UC Riverside. B. Schmookler supervised the work of the aforementioned students, planned the beam test, performed measurements, and carried out simulations; he also edited the manuscript. All of the above participated in the experimental campaign. M. Arratia conceived and supervised this work and edited the manuscript.

\section*{Acknowledgments}
We thank the ePIC Collaboration for their valuable feedback during the preliminary phases of this work. In particular, we thank Alexander Bazilevsky and the EIC project for their support in acquiring beam time for this study. We also appreciate the advice and support of the members of the California EIC Consortium, especially Oleg Tsai. We thank Sebouh Paul for designing the Hexagon Board used in this study. Finally, we are grateful to Michael Backfish, Eric J Prebys and the entire UC Davis Cyclotron staff for their help during beam operation.

We acknowledge support from the MRPI program of the University of California Office of the President (award number 00010100). Sean Preins was supported by a HEPCAT fellowship through DOE award DE-SC0022313.
\FloatBarrier
\renewcommand\refname{Bibliography}
\bibliographystyle{utphys} 
\bibliography{references.bib} 
\FloatBarrier
\newpage
\section{Appendix}
\label{Appendix}

\begin{figure}[H]
    \centering
    \includegraphics[width=0.85\linewidth]{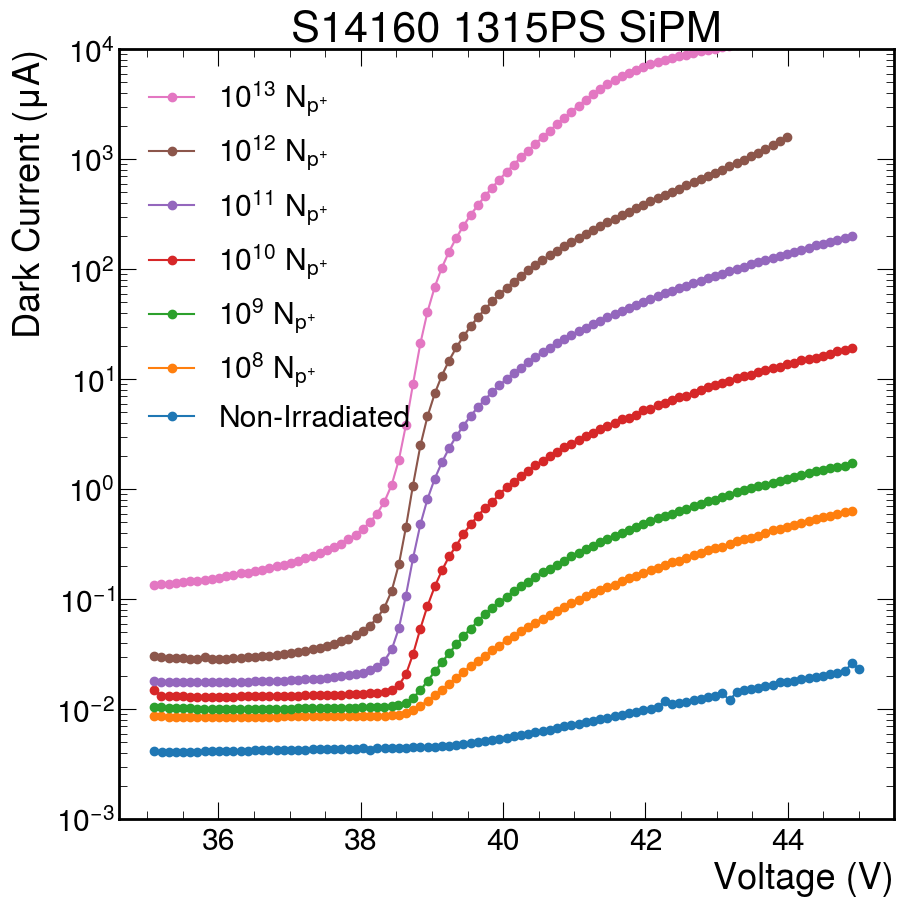}
    \label{fig:1315}
    \caption{Dark current versus applied voltage for various fluences for the Hamamatsu S14160-1315PS SiPM model. Measurements were performed at room temperature. The data include a correction factor to account for the corresponding room-temperature annealing time, as described in the main text. Each curve represents a different SiPM unit, which may exhibit slightly different overvoltages due to typical variations in this parameter.}
\end{figure}

\begin{figure}[H]
    \centering
     \includegraphics[width=0.85\linewidth]{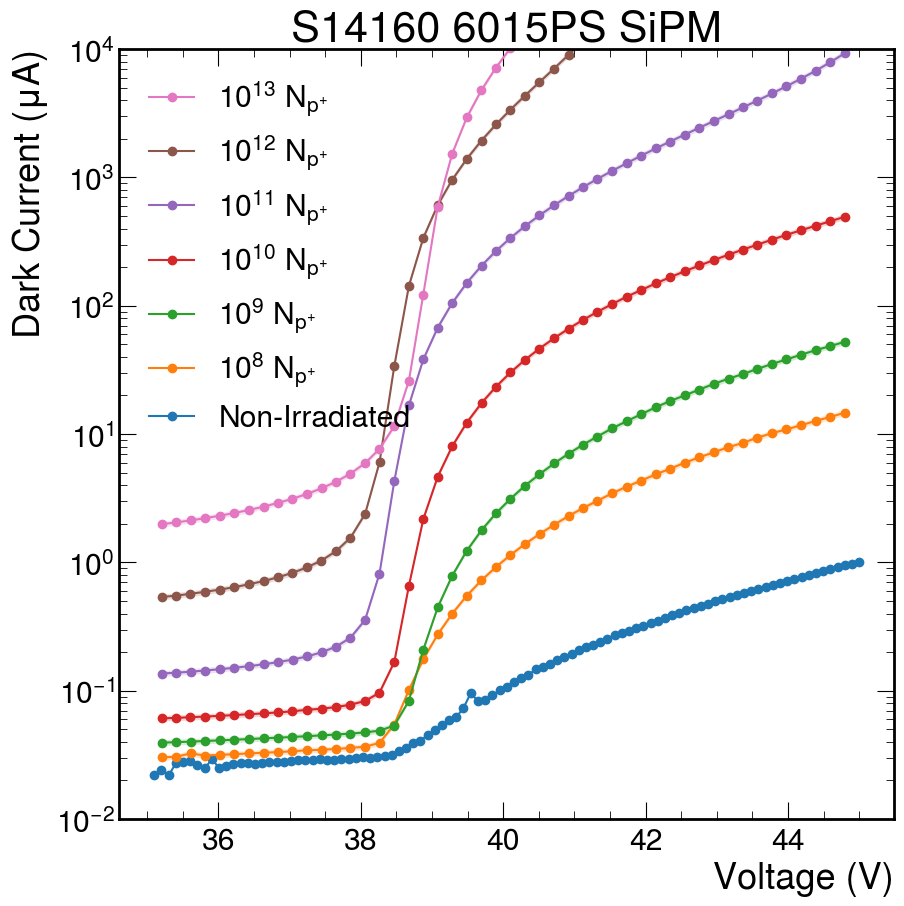} 
    \label{fig:6015}
    \caption{Dark current versus applied voltage for various fluences for the Hamamatsu S14160-6015PS SiPM model. Measurements were performed at room temperature. The data include a correction factor to account for the corresponding room-temperature annealing time, as described in the main text. Each curve represents a different SiPM unit, which may exhibit slightly different overvoltages due to typical variations in this parameter.}
\end{figure}

\begin{figure}[H]
    \centering
    \includegraphics[width=0.85\linewidth]{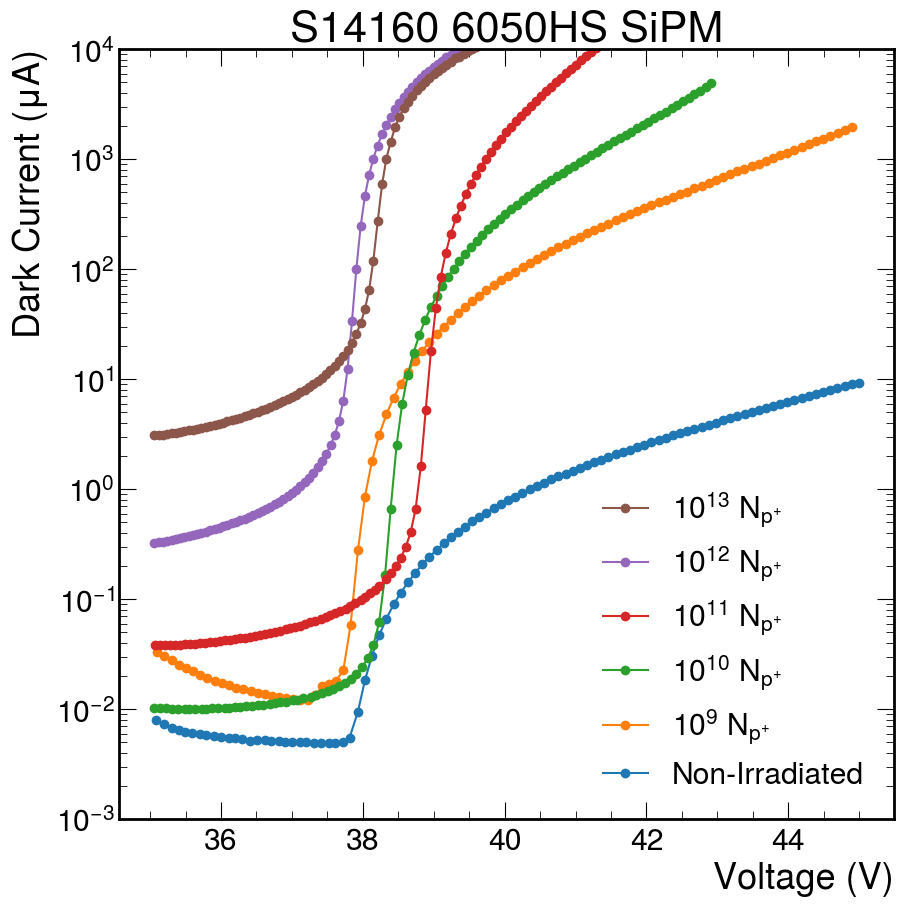} 
    \label{fig:A6050hs}%
    \caption{Dark current versus applied voltage for various fluences for the Hamamatsu S14160-6050HS SiPM model. Measurements were performed at room temperature. The data include a correction factor to account for the corresponding room-temperature annealing time, as described in the main text. Each curve represents a different SiPM unit, which may exhibit slightly different overvoltages due to typical variations in this parameter.}
\end{figure}

\begin{figure}[H]
    \centering
     \includegraphics[width=0.85\linewidth]{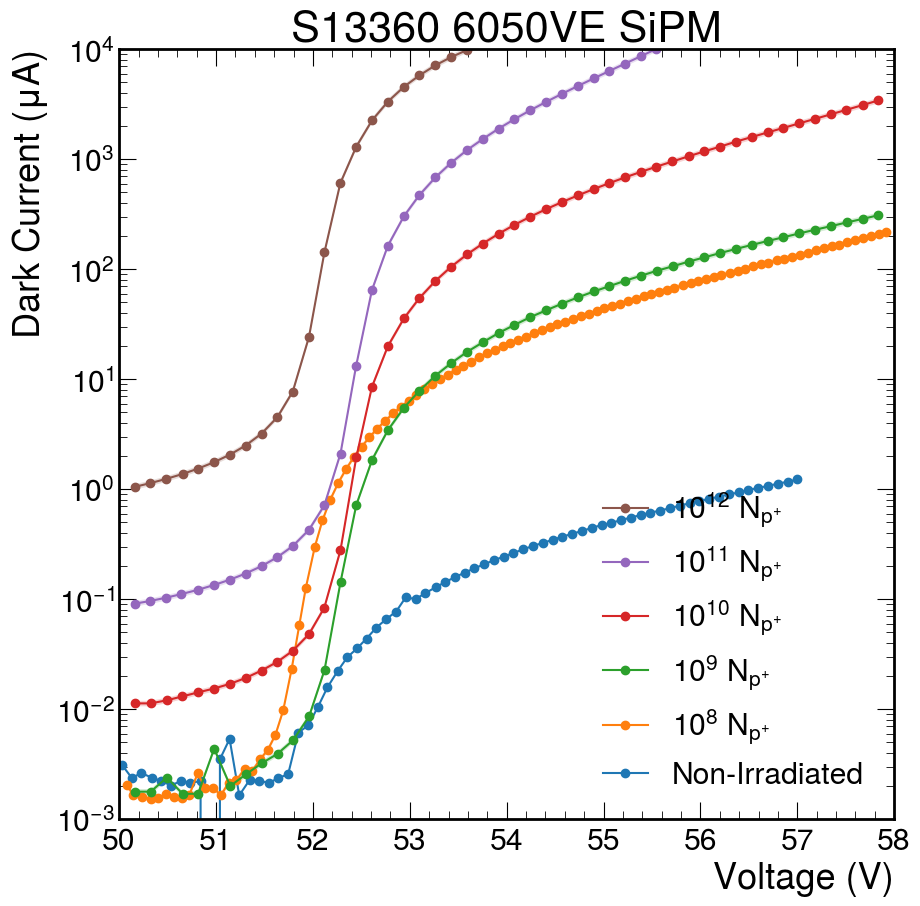} 
    \label{fig:A6050ve}%
    \caption{Dark current versus applied voltage for various fluences for the Hamamatsu S13360-6050VE SiPM model. Measurements were performed at room temperature. The data include a correction factor to account for the corresponding room-temperature annealing time, as described in the main text. Each curve represents a different SiPM unit, which may exhibit slightly different overvoltages due to typical variations in this parameter.}
\end{figure}

\end{document}